\documentclass[prc,twocolumn,showpacs]{revtex4}
\usepackage{graphicx}
\usepackage{bm}

\begin{document}

\title{Neutrino mean free path in neutron stars}

\author{Caiwan Shen$^{1,2}$, U. Lombardo$^{1,3}$,
        N. Van Giai$^{4}$, and W. Zuo$^{5}$}
 \affiliation{
   $^{1}$INFN-LNS, Via Santa Sofia 44, I-95123 Catania, Italy \\
   $^{2}$China Institute of Atomic Energy, P.O.Box 275(18),
         Beijing 102413, China \\
   $^{3}$Dipartimento di Fisica, Via Santa Sofia 64, I-95123 Catania,
         Italy \\
   $^{4}$Institut de Physique Nucl\'eaire, F-91406 Orsay,
         France \\
   $^{5}$Institute of Modern Physics, Chinese Academy
         of Sciences, Lanzhou, China }

\begin{abstract}
The neutrino propagation in neutron stars is studied in the
framework of the linear response method. The medium effects are
treated in the non-relativistic Brueckner-Hartree-Fock approach
either in the mean-field approximation or in the RPA. The residual
interaction is expressed in terms of the Landau parameters
extracted from the equation of state of spin- and
isospin-polarized nuclear matter. The Brueckner theory including
three-body forces is used for determining the equation of state.
Numerical predictions for the response function of nuclear matter
in $\beta$-equilibrium and the neutrino mean free path are
presented in a range of baryonic densities and temperatures. The
main results are a dominance of the charge-exchange component over
the scattering component and an enhancement of the neutrino mean
free path induced by nuclear correlations.
\end{abstract}


\maketitle
\section{Introduction}
The interaction of neutrinos with baryons has been mostly studied
in connection with the stability of nuclei, but it also plays a
crucial role in the thermal evolution of supernovae and
protoneutron stars, where the nuclear medium exhibits  quite
distinctive physical features related to density, temperature and
chemical composition. A large effort has been devoted to study the
production of neutrinos via direct or modified URCA processes
\cite{LATT} and, more recently, via the bremsstrahlung of nucleons
in the strong magnetic field of neutron stars\cite{SEDR}. In the
past simulation codes have incorporated the propagation of
neutrinos in neutron matter mostly by using the Fermi-gas model,
whereas the effects of correlations did not receive the due
consideration except in a few cases \cite{IWA,BURR0,BURR}. In the
recent years calculations of correlation effects have been
performed, including  phenomenological approaches with Skyrme
forces\cite{REDD,PRAK,MARG}, microscopic Brueckner theory
\cite{BURR0,BURR,SHEN,Marg03} and relativistic mean field theory
\cite{MATE,YAMA,MORN,LEIN,MORN1,HORO,GRAC,NIEM}. On the other
hand, in order to reliably describe the correlations of nuclear
matter in extreme conditions, one needs a well developed
microscopic many-body theory, since not enough experimental
constraints exist so far in a so wide range of density,
temperature and isospin asymmetry which are supposed to occur in
the present neutron-star models.

In this paper we present a study of the nuclear response function
to weak interaction and the effects of short- and long-range
correlations on the neutrino transport in neutron stars. It is
based on the equation of state (EOS) predicted by the Brueckner
theory approach including
relativistic effects and
nucleonic resonances. Recently
this approach has made a step forward in reproducing the empirical
saturation properties of nuclear matter\cite{UMB,ZUO}. The
particle-hole NN residual interaction is extracted from the Brueckner
theory and cast in terms of the Landau parameters. The response
function to the neutrino propagation is calculated first in the
Brueckner-Hartree-Fock (BHF) limit, and then in
the random phase approximation (RPA) in neutron matter in
$\beta$-equilibrium with protons and electrons for various
conditions of neutron density and temperature. The neutrino mean
free path (MFP) is also calculated taking into account the ($\nu_e, \nu_e'$)
scattering from nucleons and the ($\nu_e, e^-$) charge-exchange process. In
this work we discuss only the $\nu_e$ case which is the most important one
for the MFP issue.

\section{Nuclear matter in the BHF approximation}

A neutron star, in the most simplified model, is made of neutrons,
protons and electrons, and the relative abundance is controlled by
the $\beta$-equilibrium condition (the effect of muons is
negligible). Under the condition of charge neutrality, the proton
fraction $Y_p=Z/A$ is driven by the value of the symmetry energy
at a given total baryonic density. Here, we will always consider
nuclear matter in $\beta$-equilibrium unless explicitly
stated.

Our framework is the Brueckner-Bethe-Goldstone (BBG) approach,
where the perturbative expansion of the total energy per particle
$E_A$ can be cast according to the number of hole lines. The
lowest order defines the Brueckner-Hartree-Fock approximation,
which exhibits a satisfactory convergence provided the continuous
choice for the mean field is adopted \cite{SONG}.

In the recent years a remarkable step forward in the reproduction of
the empirical saturation properties has been performed including
three-body forces (3BF) in the BBG theory \cite{UMB,ZUO}. Their
effect in fact is twofold: on one hand, it brings the saturation
density of nuclear matter close to the empirical value; on the
other hand, it provides the high-density strong repulsive
components. As a consequence the BHF approximation has reached the
level of a consistent description of a strongly correlated Fermi
system in the region of saturation density as well as in high
density regions.

 A microscopic derivation of the full set of
$l=0$ Landau parameters can be done, based on the BHF
approximation, from the calculation of the energy per particle of
nuclear matter in different states of spin and isospin
polarization. Starting from unpolarized symmetric nuclear matter
at a given density $\rho$, we may extend the calculation of $E_A$
to isospin density variations $\delta\rho_{\tau}=
(\rho_{\tau}-\rho)/\rho$  and spin-isospin density variations
$\delta\sigma_{\tau}=
(\rho^\uparrow_{\tau}-\rho^\downarrow_{\tau})/\rho_{\tau}$. The
response of the system is related to the second derivatives of the
expansion of $E_A$,
\begin{eqnarray}
E_A(\rho,\delta\rho_n,\delta\rho_p) &=& E_A(\rho)+ \frac{1}{2}
\sum_{\tau\tau'} \Phi_{\tau,\tau'}\delta\rho_{\tau}
\delta\rho_{\tau'},   \\
E_A(\rho,\delta\sigma_n,\delta\sigma_p)&=& E_A(\rho) +\frac{1}{2}
\sum_{\tau\tau'} \Gamma_{\tau,\tau'}
\delta\sigma_{\tau}\delta\sigma_{\tau'}.
\end{eqnarray}
These derivatives give the $l=0$ Landau parameters of the residual NN
interaction,
\begin{eqnarray}
F_0 &=& F_{nn}^{0} + F_{np}^{0} = \frac{4N(0)}{\rho}(\Phi_{nn}+\Phi_{np}) - 1, \\
F'_0 &=& F_{nn}^{0} - F_{np}^{0} =
\frac{4N(0)}{\rho}(\Phi_{nn}-\Phi_{np}) - 1,\\
G_0 &=& G_{nn}^{0} + G_{np}^{0} = \frac{4N(0)}{\rho}(\Gamma_{nn}+\Gamma_{np}) - 1, \\
G'_0 &=& G_{nn}^{0} - G_{np}^{0} =
\frac{4N(0)}{\rho}(\Gamma_{nn}-\Gamma_{np}) - 1,
\end{eqnarray}
where $N(0)$ is the level density at the Fermi surface. Notice
that our definitions Eqs.(3)-(6) differ by a factor of 2 from that
of other authors, for instance \cite{SAK}. The Landau parameters
are constrained by robust physical observables. $F_0$ is in fact
constrained by the compression modulus, whose value should range
in the interval 230-250 MeV according to the experimental
determination of the monopole giant resonance energy centroid.
$F_0'$ is related to the symmetry energy, which at the saturation
density is reported about 30 MeV in the Bethe-Weiszacker mass
formula. $G_0$ is related to the spin modes, which actually are
hardly observed in nuclei. So far experimental information on
$G_0$ is not enough since spin resonances have only been observed
with too small strength compared to other collective
modes\cite{OSTE}. Finally $G_0'$ is constrained by the
Gamow-Teller giant resonance. A value of 1.2 at the saturation
point has been determined with high precision from the
experimental excitation energy of the Gamow-Teller resonance on
$^{90}$Ni \cite{SAK}. The BHF prediction of the Landau parameters
is reported in Fig.1.
\begin{figure}
\centering{\includegraphics[width=8cm,angle=0]{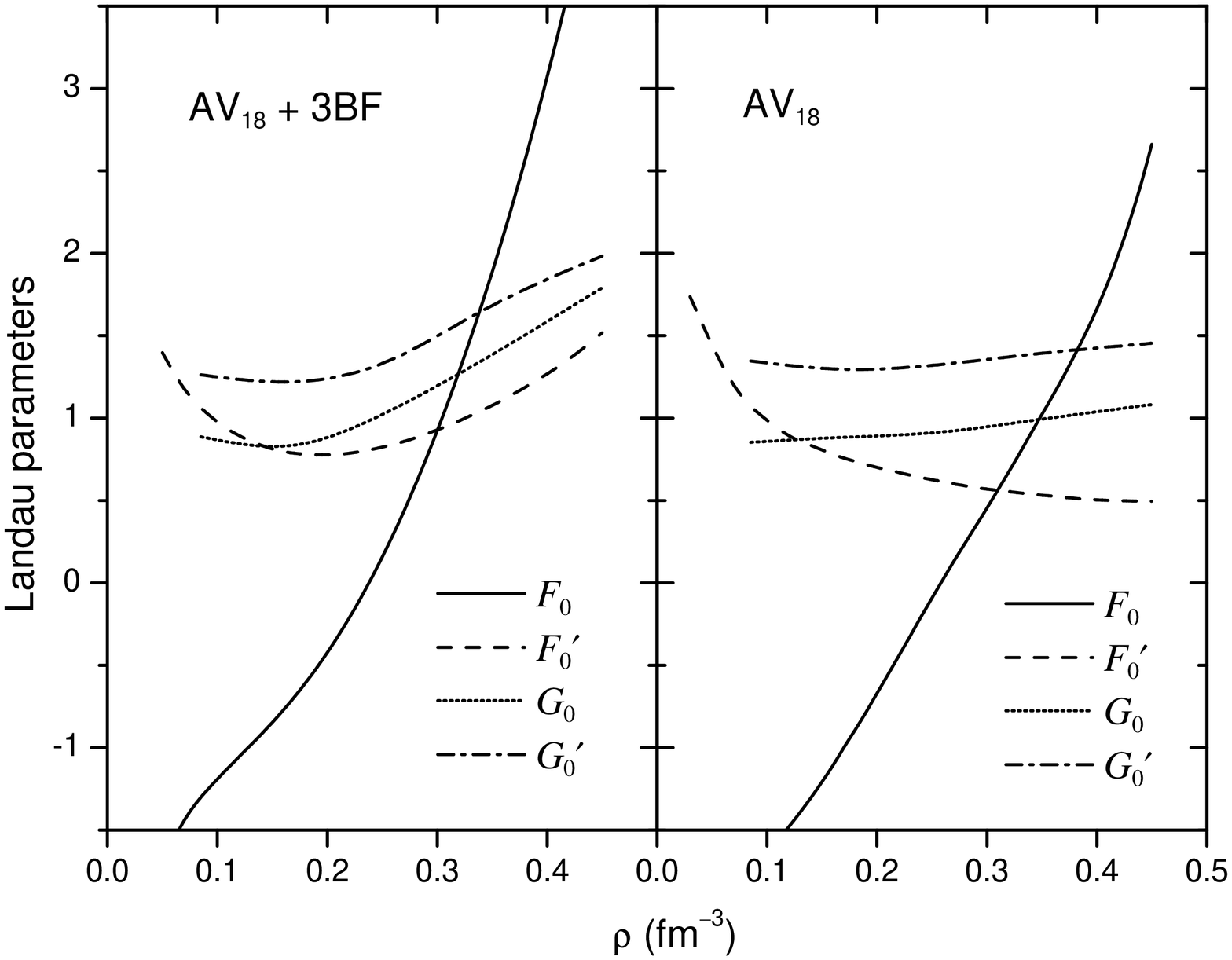}}
\caption{Landau parameters of symmetric nuclear matter with 2BF
$AV_{18}$ (right) and the same plus 3BF (left)  }
\label{fig-1}
\end{figure}
They have been obtained from a BHF calculation with the Argonne
$AV_{18}$ \cite{WIRI} as two-body force (2BF) and a microscopic three-body
force which corrects the BHF approximation by relativistic effects and
nucleonic virtual excitations\cite{GRAN}.

 The prediction of the Landau parameters
for densities other than the nuclear density is of great interest
in the study of neutron stars. In connection with the strong
magnetic fields observed in neutron stars some authors have studied the
magnetic susceptibility in neutron matter and found that
$G_0$ reduces the susceptibility of the degenerate neutron
gas\cite{FANT,VIDA,FRASC,BOMB} . This reduction is amplified at
high density when including 3BF either in Brueckner
calculations\cite{FRASC} or in Monte Carlo many-body simulations
\cite{FANT}.

In this work it is  assumed that the Landau parameters do not
change appreciably with temperature, which is certainly a good
approximation below the liquid-to-gas phase transition
($T_c\approx 18 MeV$). This point will be discussed in more
details in section IV(A).

\section{Interaction of neutrinos with matter}

During their propagation in neutron-star matter neutrinos
experience collisions with nucleons via weak coupling with the
nucleon neutral currents
$j_{\mu}=\overline{\psi}_{\tau}\gamma_{\mu}(c^{\tau}_v
-c^{\tau}_a\gamma_5)\psi_{\tau}$. Neutrinos can also disappear in
the charge-exchange process $\nu_e + n \rightarrow e^- + p$ and we
also consider the coupling with charged currents
$j_{\mu}=\overline{\psi}_{p}\gamma_{\mu}(g_v-g_a\gamma_5)\psi_{n}$
giving rise to neutrino absorption. The vector and axial coupling
constants for neutral currents are $c^{\tau}_v$ and $c^{\tau}_a$,
and for charged currents $g_v$ and $g_a$ (a complete list of the
constants is given in Ref.\cite{PRAK}).

The neutrino MFP $\lambda$ is derived from the
transport equation
\begin{equation}
\frac{c}{\lambda_{\vec k}} = \sum_{\vec k'} W_{fi}(k-k')(1-n_{\vec
k'})+ W_{fi}(k'-k) n_{\vec k'},
\end{equation}
where $n_{\vec k}$ is the occupation number of neutrinos and
$W_{fi}$ is the transition probability corresponding to the individual
processes under consideration.

 In the non-relativistic limit the transition rate can be
 written\cite{BURR0,BURR,IWA}
\begin{equation}
W_{fi}= G^2_w [(1+\cos\theta)W_v + (3-\cos\theta)W_a],
\end{equation}
 where $G_w$ is the weak coupling constant, $\theta$ is the scattering angle and $W_v$ ($W_a$) is the
 vector (axial) term.
The transition rates are expressed in terms of the nuclear
structure functions $S_{v,a}^{\tau\tau'}$. In the case of nucleon
scattering the relation is
 \begin{eqnarray}
 W_{v,a} &=& \sum_{\tau,\tau'} c_{v,a}^{\tau} c_{v,a}^{\tau'}
 S_{v,a}^{\tau\tau'},\\
S_{v,a}^{\tau\tau'} &=& -\frac{1}{\pi}\frac{1}{1-e^{q_0/kT}}{\rm
Im} \chi_{\tau,\tau'}^{v,a}
\end{eqnarray}
For the charge-exchange process the relation is
\begin{eqnarray}
W_{v,a} &=& g^2_{v,a} S_{v,a}^{pn}, \\
S_{v,a}^{\tau\tau'} &=&
-\frac{1}{\pi}\frac{1}{1-e^{(q_0+\delta\mu)/kT}}{\rm Im}
\chi_{p,n}^{v,a}.
\end{eqnarray}
The quantity $q_0$ is the energy transfer and $\delta\mu$ is the
shift between the neutron and proton chemical potentials. From the
previous equations we see that the structure functions are, in
turn, related to the response functions $\chi$ which will be discussed
in the next section.

\section{Nuclear Response functions}

The major part of the work needed to obtain the neutrino MFP is in the
calculation of the structure functions.
These calculations can be performed at different levels of
approximation beyond the free Fermi gas model. We will consider
first the BHF approximation, which already takes into account the
strong short-range correlations via the mass renormalization and
the depletion of the Fermi surface; second, the particle-hole
interaction will be taken into account in the RPA approximation.
Since the ring diagram summation is  almost a prohibitive task
with the Brueckner G-matrix, we may approximate, in the
low-frequency limit ${\omega}/({kv_F}) \ll 1$, the G-matrix
residual interaction by a Landau-Migdal form
whose parameters are extracted from the BHF theory
as described in the previous section.

\subsection{BHF response function}

In the BBG approach the expansion of the self-energy
$\Sigma_{\tau}(p,\omega)$can also be cast as a hole line expansion,
but the lowest order in the self-energy expansion  does not
correspond to the lowest order in the expansion of the energy per
particle. According to the Landau definition of the quasiparticle
energy, the BHF approximation for the self-energy should include
the expansion up to the second order, the latter being the so
called rearrangement term \cite{EBHF}. Doing so, the
Hugenholtz-Van Hove theorem is fulfilled. Once the approximation
for $\Sigma$ has been settled, the single-particle propagator
takes the form
\begin{equation}
G_{\tau}^{-1}(p,\omega) = \omega - \frac{p^2}{2m} -
\Sigma_{\tau}(p,\omega) + e^{\tau}_F \approx
\frac{1}{Z^{\tau}_p}(\omega - \epsilon_p^{\tau}),
\end{equation}
where $\epsilon^{\tau}_p$ is the single-particle energy
\begin{equation}\label{eps}
\epsilon_p^{\tau} = \frac{p^2}{2m}+\Sigma(\epsilon_p)-e^{\tau}_F
\end{equation}
and $\epsilon^{\tau}_F$ is the Fermi energy. The factor
$Z^{\tau}_p =
    (1-\frac{\partial\Sigma^{\tau}}{\partial\omega})_{\omega=\epsilon_p}^{-1}
$ is the quasiparticle strength associated to the depletion of the
occupation probability of the single-particle level $\epsilon_p$.
In the present study the rearrangement term will be neglected,
which is not a too severe approximation since it does not
affect the EOS at zero temperature and, consequently, the
calculation of the Landau parameters . At finite temperature it
cannot be absolutely neglected, since the single-particle spectrum
determines the thermal Fermi functions and hence the EOS itself.

The neutrino propagation is studied in the framework of the linear
response theory. Let us first discuss the BHF limit where there
are no correlations coming from the residual interaction. The
dynamical structure functions are related to the imaginary part of the
retarded polarization or response function $\chi^{(0)}_{\tau,\tau'}(\vec
 q,q_0)=({\rm Re}+i{\thinspace\rm sign}(q_0){\rm Im})
 \thinspace\Pi^{(0)}_{\tau,\tau'}(\vec q,
q_0)$, $\Pi^{(0)}$ being the polarization function
\begin{equation}
 \Pi^{(0)}_{\tau,\tau'}(\vec q, q_0)=
-2i\int\frac{d\omega'}{2\pi}\frac{d^3q'}{(2\pi)^3}G_{\tau}(\omega',\vec
 q')G_{\tau'}(q_0+\omega',\vec q+\vec q'),
\end{equation}
where the propagators are calculated according to Eqs.(13-14). One
easily finds
\begin{equation}
\chi^{(0)}_{\tau,\tau'}(\vec q,q_0) =-\int\frac{d^3
p}{(2\pi)^3}Z^{\tau}_p Z^{\tau'}_{|\vec p+\vec q |} 
\frac{n^{\tau}_p-n^{\tau'}_{|\vec p+\vec q|}}
 {q_0-\epsilon^{\tau}_p-\epsilon^{\tau'}_{|\vec p+\vec
 q|}+i\eta}.
 \end{equation}
According to the previous discussion the thermal occupation
numbers
\begin{equation}
n_p^{\tau} = \frac{1}{1+
e^{(\epsilon^{\tau}_p-\epsilon_F^{\tau})/k_BT}}
\end{equation}
are calculated from the single-particle spectrum frozen at $T=0$.

\subsection{RPA response function }

In this  subsection we study the effect of the residual
particle-hole interaction. As already anticipated, the residual
interaction will be described in terms of the $l=0$ Landau
parameters discussed earlier. This makes it easier to calculate
the response function $\chi^{(\rm RPA)}(\vec q,q_0)$ in the RPA
limit.

Let us consider first the nucleon scattering. In this case the RPA
response function is given by the Bethe-Salpeter
equation in the 2$\times$2 isospin space
\begin{equation}
\chi_{S} = \chi^{(0)} + \chi^{(0)} L^{(S)} \chi_{S},
\end{equation}
where only the diagonal elements $\chi^{(0)}_{\tau,\tau}$ come
into play and $L^{(0)}_{\tau,\tau'}\equiv F^0_{\tau,\tau'}$ and
$L^{(1)}_{\tau,\tau'}\equiv G^0_{\tau,\tau'}$ are the matrix
elements of the particle-hole residual interaction expressed by
the Landau parameters (see Eqs.(3)-(6)). The $S=0$ ($S=1$)
component is the vector (axial) part of the response function.

In the case of charge-exchange processes the Bethe-Salpeter
equation is a one-dimensional equation
\begin{equation}
\chi_{S} = \chi^{(0)}_{p,n} + \chi^{(0)}_{p,n}  M^{(S)}  \chi_{S},
\end{equation}
where $M^{(0)}\equiv F'_0$,  $M^{(1)}\equiv G'_0$, and
$\chi^{(0)}_{pn}$ is the off-diagonal matrix element of the BHF
response function.

The structure functions are essentially given by the imaginary
part of the response functions, according to Eqs.(10) and (12).

\section{Results and discussion}%

The input data, namely the BHF mean field for the $\chi^{(0)}$
response functions and the Landau parameters for the RPA response
functions, are obtained from a non-relativistic BHF calculation of
nuclear matter with the two-body $AV_{18}$\cite{WIRI} and the
microscopic 3BF \cite{GRAN}. The temperature is set to zero, which
is actually a good approximation only  below the critical point,
as already mentioned. The proton fraction is calculated from the
symmetry energy extracted in the calculation of isospin-asymmetric
nuclear matter. For $\rho=0.34$ fm$^{-3}$, $Y_p=0.077$ without
the 3BF and $Y_p=0.167$ with the 3BF. Thus, the 3BF has a sizable
effect on $Y_p$.

The unperturbed (BHF) response function, as described by Eq.(16),
is plotted in Fig. 2 for different temperatures and different
processes. As it can be seen in the left panels (imaginary parts),
the $|\chi^{(0)}($nn$^{-1})|$ is always greater than
$|\chi^{(0)}($pp$^{-1})|$. This effect, which arises from the
small proton fraction, i.e., large asymmetry parameter
$$
\beta\equiv \frac{N-Z}{A}
$$
in neutron stars and smaller
$Z_{k}^{\text{p}}$ compared to $Z_{k}^{\text{n}}$ from BHF
calculations, gives larger scattering probability for $\nu+n
\rightarrow \nu+n$ and, consequently, shorter neutrino MFP. The
response function related to scattering contributes to the
neutrino MFP essentially for $q_0\ge 0$ whereas the one related to absorption
contributes for $q_0\ge(\mu_p-\mu_n)$.
 It also shows that
the response function only depends softly on the temperature, at
least in the $T=0$ Brueckner approximation, since the temperature
appears
only through $n_{k}^{\tau}$.
\begin{figure}
[tbh]
\begin{center}
\includegraphics[width=8.7cm]{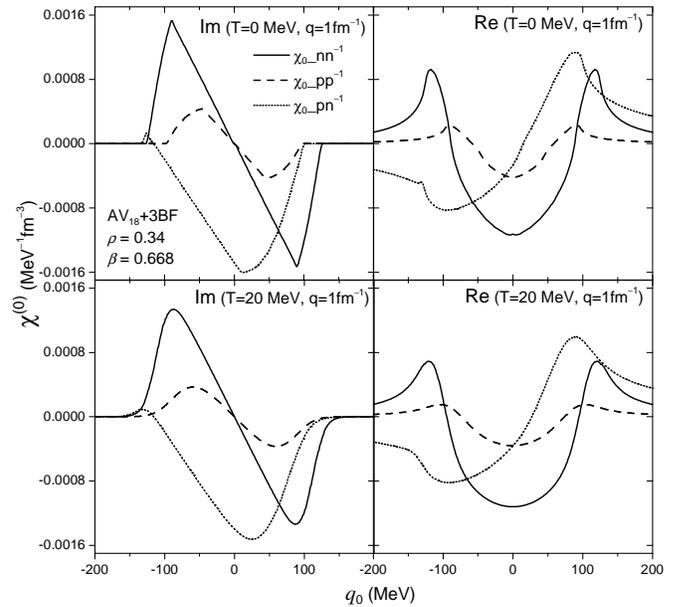}
\caption{The uncorrelated response functions for $(\nu, \nu')$
scattering on neutrons (nn$^{-1}$, solid), on protons (pp$^{-1}$,
dash) and neutrino absorption (pn$^{-1}$, dot).The imaginary part
(left panel) and the real part (right panel) are plotted as
functions of energy transfer. The $\rho$, $\beta$ and momentum
transfer $q$ are fixed at 0.34 fm$^{-3}$, 0.668, 1.0 fm$^{-1}$,
respectively. The upper and lower panels correspond to $T$=0 and
$T$=20 MeV.}
\label{Fig-2}%
\end{center}
\end{figure}

Fig.3 shows the imaginary part of the RPA response function
as calculated from the Landau parameters with 3BF included.
 Here, we
show
the vector (left side) and axial (right side)
components for the three isospin channels in consideration.
The comparison emphasizes the dominance in both $S=0$ and $S=1$
channels of the absorption component (see insets). This effect is
due to the $q_0$ threshold (compare Eq.(10) and Eq.(12))
associated with the
charge-exchange process as already discussed above.

The resonant structure of all curves is due to the collective
excitations of the medium: zero sound in the case of the vector
component and spin waves in protons and neutrons in the case of
the axial component.

\begin{figure}[tbh]
\begin{center}
\includegraphics[width=8.7cm]{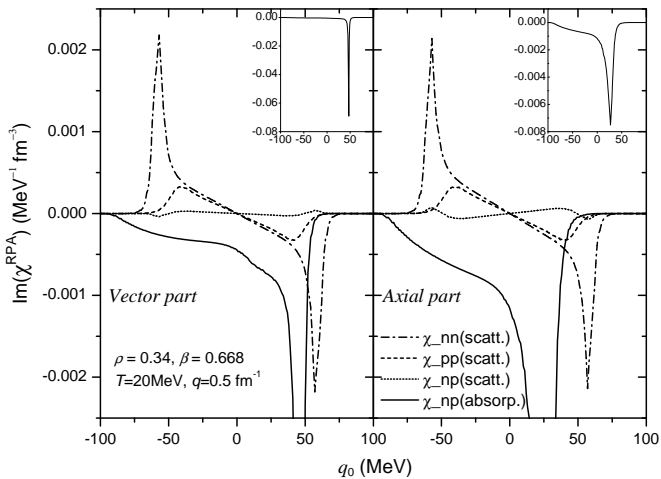}
\caption{Imaginary parts of RPA response functions.  The vector
components (left panel) and the axial-vector components (right
panel) are plotted as a function of energy transfer for fixed
nucleonic density, temperature and momentum transfer.}
\label{Fig-3}%
\end{center}
\end{figure}

 With the above response functions the neutrino MFP can be
evaluated. The calculated MFP are shown as functions of the density
in Fig.4
for neutral current process on the left side and charged current
process on the right side. $T$ and $E_{\nu}$ are fixed to 20 MeV and
40 MeV, respectively, and $\beta$ is fixed according to $\beta
$-equilibrium for each density and different approximations. In
order to see the different effects from the scattering process and
absorption process, we divide the neutrino MFP
into two parts: $\lambda_{\text{scatt.}}$ for scattering process
(left panel) and $\lambda_{\text{absorp.}}$ for neutrino
absorption process (right panel). Then, the final neutrino MFP is
$\lambda$ = $(\lambda_{\text{scatt.}}^{-1}+\lambda_{\text{absorp.}}^{-1}%
)^{-1}$. Comparing the left and right panels, we get the first
message that the neutrino MFP caused by the neutrino absorption is
much smaller than that caused by neutrino scattering. This is a
well-known effect which is due to two reasons, as noticed by
Reddy et al.\cite{REDD}. First, the absorption transition
probability is four times the scattering transition probability;
second, the absorption rate is controlled by the electron chemical
potential while the scattering is controlled by the neutrino
chemical potential. The inclusion of RPA correlations in the
structure functions has the effect of reducing the phase space
available for scattering and the MFP turns out to be substantially increased
compared to the free Fermi-gas model and also to the BHF
model, the latter only at high density. According to
Ref.\cite{ZUO}, the proton fraction in neutron stars increases
dramatically
at higher densities when the
3BF is included. This affects the behavior of the neutrino MFP in
both scattering and absorption processes: the neutrino MFP
increases with increasing density, especially for the neutrino
absorption process. This is related to the fact that with
decreasing $\beta$, the proton Fermi energy increases and then the
reaction $\nu_e + n\rightarrow e^{-}+p$ has a smaller cross section.

\begin{figure}
[ptb]
\begin{center}
\includegraphics[ width=8.7cm]{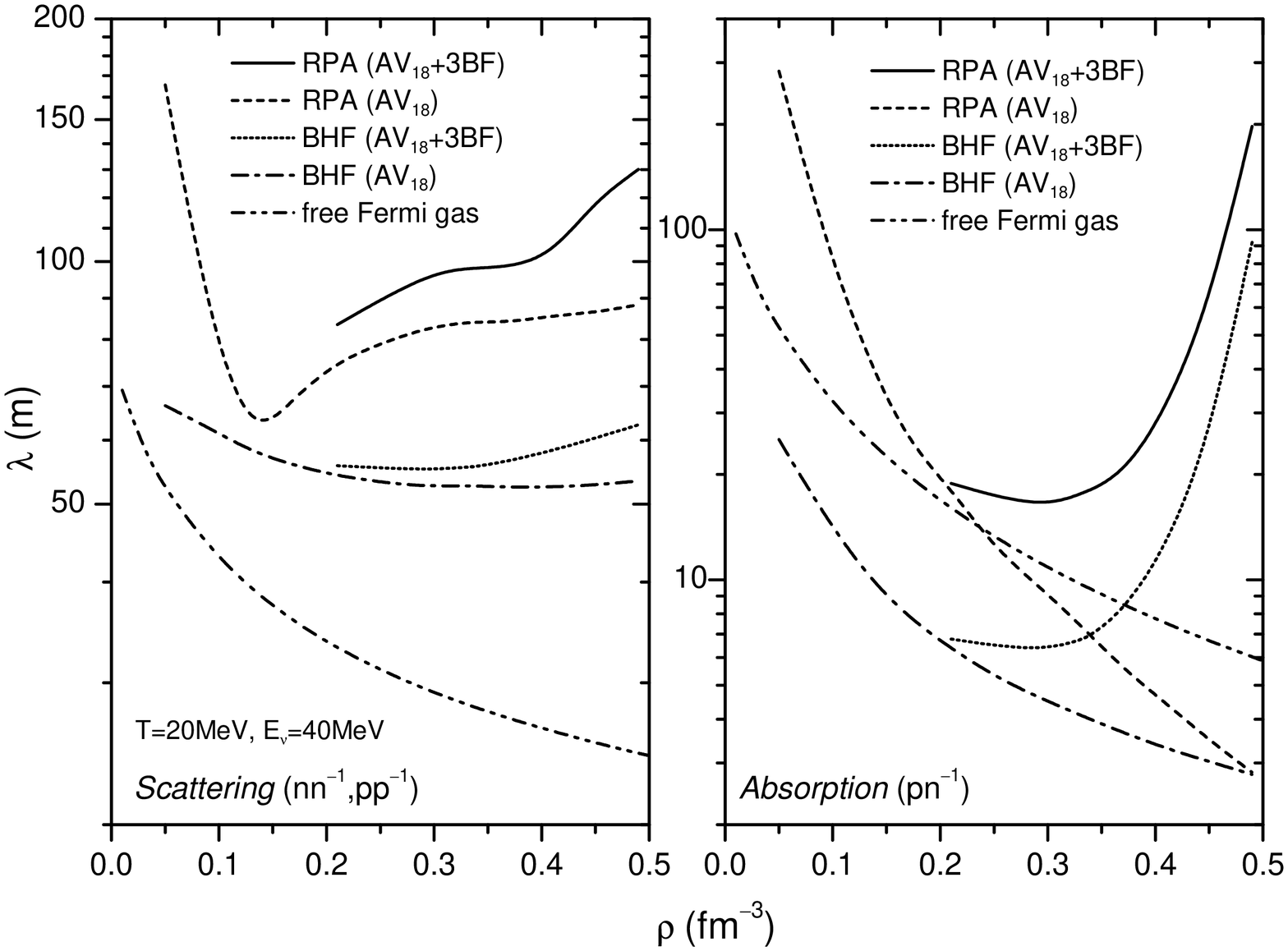}
\caption{Neutrino MFP for scattering (left
panel) and absorption (right panel). Solid (dashed) lines:
RPA with (without) 3BF; dotted (dash-dotted) lines:
BHF with (without) 3BF; dash-dot-dotted
lines: free Fermi-gas model. The temperature $T$ and
neutrino energy $E_{\nu}$ are fixed to 20 MeV and 40
MeV, respectively.}%
\label{fig-4}%
\end{center}
\end{figure}

Since the temperature in the medium drops out rapidly in the
protoneutron star, it is interesting to show the relation between
neutrino MFP and temperature. In Fig. 5 (left) we plot the
neutrino MFP as a function of temperature with $\rho$, $\beta$ and
$E_{\nu}$ fixed at 0.34 fm$^{-3}$, 0.668 and 40 MeV, respectively.
One can see the different behaviour of the scattering and
absorption MFP when $T$ increases. For the scattering process,
$\lambda$ decreases by a factor 30 from $T$=5 MeV to $T=60$ MeV
while for the neutrino absorption process this factor is only 2.5
. In the right side of Fig. 5 the $E_\nu$ dependence is also
depicted. We just notice that the ratio between scattering and
absorption MFP is almost constant.
\begin{figure}
[tbh]
\begin{center}
\includegraphics[width=8.7cm]{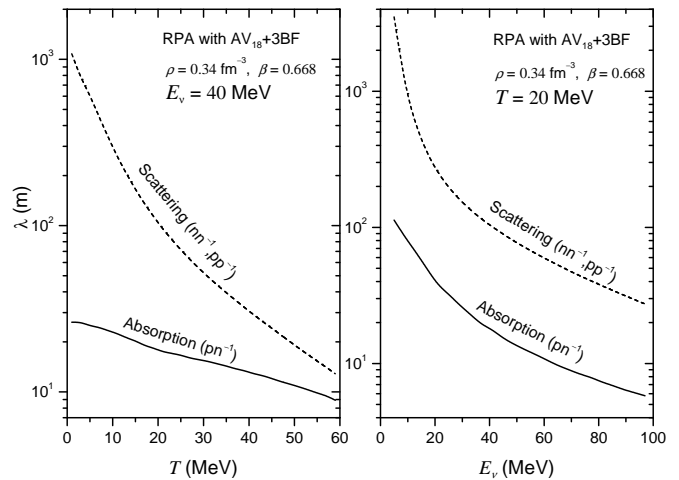}
\caption{Neutrino MFP vs. temperature (left) and vs. neutrino
energy (right). The solid line is for neutrino absorption and
dashed line for neutrino scattering. The values of $\rho$,
$\beta$,
are 0.34 fm$^{-3}$, 0.668, respectively.}%
\label{fig-5}%
\end{center}
\end{figure}

\section{Comparison with other approaches and conclusions}%

The nuclear response function and the neutrino MFP in $\beta$-
equilibrium nuclear matter have been calculated on the one hand in the BHF
mean field approximation, and on the other hand in the RPA using a
Landau-Migdal residual interaction to evaluate the effects of long-range
correlations.
Mean fields and residual interaction
have been obtained within the non-relativistic
Brueckner-Hartree-Fock theory. The two-body NN potential has been
supplemented by a microscopic three-body potential, which  takes
into account relativistic effects and also nucleonic excitations.
From the present calculations one may conclude that the reduction
of the neutrino MFP in the limit of the BHF mean field is much
less pronounced than in the Fermi-gas model when going from low to
high density .This reveals the important role of the short-range
correlations incorporated in the BHF nucleon propagators. In
addition, the effects of the residual particle-hole interaction
have also a deep influence on the neutrino transport in
$\beta$-equilibrium nuclear matter, resulting in a sizeable
enhancement of the neutrino MFP. This result confirms the
conclusion of a previous calculation of neutrino MFP in pure
neutron matter \cite{SHEN}. At high density this effect is
magnified in the absorption channel by the intense repulsive
three-body force.

Since the interaction is poorly known at densities of interest in
neutron star physics, the comparison with other approaches is
difficult and leads often to contradictory predictions. Among the
non-relativistic calculations we should quote the pioneering
results by Burrows and Sawyer\cite{BURR0,BURR}. In spite of using a
different set of Landau parameters (in particular $G_0 = 0$)
the main features of neutrino MFP they predict are in good
agreement with ours. In particular, they find an increase of $\lambda_{\nu}$
by up to one
order of magnitude with increasing density as an
overall effect of medium correlations, and a similar temperature
dependence. In the recent work of Margueron
et al.\cite{Marg03} the emphasis is put on temperature effects in a
calculation of MFP based on BHF and RPA but the 3BF is not treated and the
important contribution to the MFP due to neutrino absorption is not
calculated. It is also found that the scattering MFP including RPA
correlations does increase with density, similarly to our present results.

Non relativistic phenomenological approaches
\cite{REDD,PRAK, MARG} manifest an opposite
 trend: the RPA neutrino MFP is quenched by the density increase. In fact, the
residual interaction in the spin channel instead of slowly
increasing at high density, decreases and finally becomes
attractive. This explains also the prediction of a ferromagnetic
transition in phenomenological approaches.

Relativistic approaches lead to an enhancement of the neutrino MFP
when including the interaction effects in RPA even if some
disagreement still exists among different calculations. In one case
\cite{MORN} the reduction of the scattering rate is estimated about
10$\%$ to 15$\%$, in others \cite{REDD,YAMA} the effect is much
stronger.

A general conclusion is that the neutrino MFP is very sensitive to
the EOS of nuclear matter in all aspects and it must be calculated
as much as possible consistently with the EOS. In this respect we
should mention a serious drawback of most calculations, which is
the temperature dependence of the effective interaction. A typical
value of temperature of about 20 MeV maybe is large for neglecting
the effects of thermal excitations of the medium on the in-medium
interaction, for instance the G-matrix in the Brueckner approach.

In conclusion, our calculation predicts a quenching of the opacity
of nuclear matter to the neutrino propagation and then provides the
quantitative basis to estimate the energy release in the
supernovae explosions and the rapid cooling of a protoneutron star.

\end{document}